\documentclass[aps,prl,reprint,showpacs,showkeys,superscriptaddress,preprintnumbers,groupedaddress,nofootinbib]{revtex4-1}
\usepackage{amsmath,amssymb,bm,mathrsfs}
\usepackage{epsfig}
\usepackage{slashed}
\usepackage{xcolor}
\usepackage{subfig}
\usepackage{srcltx}
\usepackage{multirow}
\usepackage[colorlinks=true,linkcolor=blue,citecolor=blue,urlcolor=violet]{hyperref}

\newcommand{\cO}{\mathcal{O}}
\newcommand{\met}{\slashed{E}_T}

\begin{document}

\title{\Large Constraining Quirky Tracks with Conventional Searches}

\author{Marco Farina}
\email{farina.phys@gmail.com}
\affiliation{
New High Energy Theory Center, Department of Physics, Rutgers University, \\
136 Frelinghuisen Road, Piscataway, NJ 08854, USA}

\author{Matthew Low}
\email{mattlow@ias.edu}
\affiliation{
School of Natural Sciences, Institute for Advanced Study, \\
Einstein Drive, Princeton, NJ 08540, USA}

\begin{abstract}
Quirks are particles that are both charged under the standard model and under a new confining group.  The quirk setup assumes there are no light flavors of the new confining group so that while the theory is in a confining phase, the distance between quirk-antiquirk pairs can be macroscopic.  In this work, we reinterpret existing collider limits, those from monojet and heavy stable charged particle searches, as limits on quirks.  Additionally, we propose a new search in the magnetic-field-less CMS data for quirks and estimate the sensitivity.  We focus on the region where the confinement scale is roughly between 1 eV and 100 eV and find mass constraints in the TeV-range, depending on the quirk's quantum numbers.
\end{abstract}

% \pacs{}
% \keywords{} 
\maketitle

%%%%%%%%%%%%%%%%%%%%%%%%%%%%%%%%%%%%%%%%%%%%%%%%%%%%%%%%%%%%%%%%%%%%%%%%%
\noindent {\bf  Introduction ---} 
The Large Hadron Collider (LHC) has now been running for several years and continues to be our most direct probe of electroweak-scale physics.  The primary directions of phenomenological studies have been naturalness-driven and signature-driven.  Along the signature-driven direction, only relatively small developments have been made in the study of unusual particle tracks.  Track reconstruction at colliders relies on the simple assumption that all particles follow simple helical trajectories characteristic of the motion of charged particles in a magnetic field.  There are new physics scenarios, however, that transcend that assumption and give rise to much stranger types of tracks at the LHC.  Some examples of these track signatures include tracks that abruptly change direction (kinked tracks), tracks that begin partway through the detector (appearing tracks), tracks with anomalous deposits of energy, and tracks with unusual curvature (see~\cite{Burdman:2008ek,Harnik:2008ax,Asai:2011wy,Harnik:2011mv,Meade:2011du,Fok:2011yc,Jung:2015boa,Knapen:2016hky} for past and related theory studies).  The latter case is typical of quirks, which will be the focus of this work.

Quirks are particles that are both charged under the standard model (SM) and under a new confining group~\cite{Okun:1980kw,Okun:1980mu,Kang:2008ea}.  The quirk setup assumes there are no light flavors of the new confining group so that while the theory is in a confining phase, the distance between quirk-antiquirk pairs can be macroscopic.  This leads to an interesting array of collider signatures based on the length, $\ell$, of the flux tube, or string, between the quirks.

It is when $\ell$ becomes comparable to the length scales relevant for detectors that quirk tracks exhibit unusual curvature.  Due to the challenges in identifying such tracks, there have been very few dedicated searches performed for quirks.  A search from DZero sets the only bound on quirks which is $m_Q~\gtrsim~120~{\rm GeV}$ when $10~{\rm nm}~\lesssim~\ell~\lesssim~100~{\rm \mu m}$ (where the individual quirks are not resolved)~\cite{Abazov:2010yb}. 

In this work, we will show for the first time that strong bounds can be set on quirks, at collider-relevant scales, using entirely standard LHC searches with no modifications to tracking algorithms.  These searches are sensitive for macroscopic string lengths.  In addition to reinterpreting existing searches, we propose a new search that can be performed in the magnetic-field-less ``0T'' data of CMS (still using standard tracking algorithms).  In the 0T data all known charged particles are expected to leave straight tracks making this dataset a nearly background-free sample for certain types of tracks with anomalous curvature.  While we propose a specific search for quirks, we are optimistic that the use of such a dataset can be extended to other scenarios beyond quirks.

The remainder of the paper is organized as follows: first, we briefly describe quirk models and their collider phenomenology.  Then, we reinterpret monojet searches and heavy stable charged particle (HSCP) searches in the quirk parameter space, leveraging the fact that hadron colliders automatically scan a range of $\ell$ largely due to the velocity distribution of particles produced.  Next, we suggest a novel use of the 0T data from CMS to constrain quirks.  Finally, we conclude with projected results and comments on the remaining quirk parameter space. \\

%%%%%%%%%%%%%%%%%%%%%%%%%%%%%%%%%%%%%%%%%%%%%%%%%%%%%%%%%%%%%%%%%%%%%%%%%%
\noindent {\bf  Quirks (at the LHC) ---} 
We now introduce the minimal ingredients for a quirk model.  To the SM gauge group we add a new ``infracolor'' gauge group that is assumed to be asymptotically-free with a confinement scale $\Lambda$ and to the SM particle content we add a new species, $Q$, with mass $m_Q$ and infracolor representation size $N_c$.  The particle $Q$ is called a quirk when it is much heavier than the confinement scale ($m_Q~\gg~\Lambda$).  Since $Q$ is assumed to be the lightest infracolored particle, there are no particles lighter than $\Lambda$ that can form ``hadrons'' and the only hadronic states are glueballs with masses a little above $\Lambda$.  When quirks are pair produced, for instance  at the LHC, there are no light hadrons to break the infracolor flux tube between the quirks.  This flux tube, or string, connecting the pair can be macroscopic in size and its tension results in an attractive force between the two particles.

If the quirks are colored they hadronize, via QCD, forming color-neutral states with SM quarks or gluons, similar to the well-studied case of $R$-hadrons in supersymmetry (we will adopt the name $R$-hadron for the color-neutral quirk-parton state).  The two $R$-hadrons are still connected by the infracolor flux tube and for all practical purposes the effect of QCD hadronization on the quirk dynamics can be neglected as $m_Q~\gg~\Lambda_{\rm QCD}$.  On the other hand, the electric charge of the $R$-hadron, does affect its detection; we will return to this point in the discussion of our results.  Clearly, tracking methods cannot be used for neutral $R$-hadrons.

We can now study the trajectory of quirks at the LHC. The equations of motion are given by the Nambu-Goto action with point masses on the ends of the string in an electromagnetic background~\cite{Kang:2008ea,Luscher:2002qv}.  The equations of motion for a single quirk are
\begin{equation} \label{eq:eqm} \begin{aligned}
\frac{\partial}{\partial t}\left( \frac{m_Q \vec{v}}{\sqrt{1 - \vec{v}_\perp^2 - \vec{v}_\parallel^2}} \right) &
\\ = - T \sqrt{1 - \vec{v}_\perp^2} \hat{s} 
& - T \frac{v_\parallel \vec{v}_\perp}{\sqrt{1 - \vec{v}_\perp^2}}
+ q \vec{v} \times \vec{B}.
\end{aligned}\end{equation}
Above $\hat{s}$ is a unit vector that points towards the other quirk and is used to define $v_\parallel = (\vec{v} \cdot \hat{s})$, $\vec{v_\parallel} = v_\parallel \hat{s}$, and $\vec{v}_\perp = \vec{v} - \vec{v}_\parallel$.  The quirk charge is $q$ and the magnetic field is $\vec{B}$.  For both the HSCP and monojet searches we use the CMS magnetic field of $\vec{B} = (0,0,3.8~{\rm T})$ while for the 0T search we use $\vec{B} = (0,0,0)$.  The tension is given by $T$ which is proportional to $\Lambda^2$.  There have been estimates that $T \simeq 1.6\Lambda^2$ in QCD~\cite{Deur:2014qfa}, but we take $T = \Lambda^2$ for simplicity (as the difference is simply a rescaling of the parameter space).

In the absence of external forces and when the quirks are back-to-back the maximum distance between them can be calculated to be
\begin{equation}
\ell_{\rm eff} = \frac{2m_Q}{\Lambda^2}(\gamma - 1) = \frac{m_Q}{\Lambda^2}v^2 + \cO(v^4),
\end{equation}
where $\gamma$ is the boost factor.  While the true string length, $\ell$, can be different, we use $\ell_{\rm eff}$ as a simple approximation.  The $m_Q/\Lambda^2$ factor follows from dimensional analysis and the $v^2$ factor plays a relevant role in collider searches.  Numerically, one has
\begin{equation} \label{eq:leff}
\ell_{\rm eff} \approx 
10~{\rm m}
\left(\frac{m_Q}{1~{\rm TeV}}\right)
\left(\frac{100~{\rm eV}}{\Lambda}\right)^2
\left(\frac{v}{0.7}\right)^2.
\end{equation}
From Eq.~\eqref{eq:leff} one can map different types of searches to the appropriate range of $\Lambda$.  For $10~{\rm keV} \lesssim \Lambda \lesssim 1~{\rm MeV}$ the quirks only separate a microscopic distance, comparable to the typical tracking resolution ($\sim{\rm \mu m}$) so that the quirk pair is observed as a single highly-ionizing straight track~\cite{Abazov:2010yb}.  For $1~{\rm eV} \lesssim \Lambda \lesssim 10~{\rm keV}$ one finds that $\ell_{\rm eff}$ is macroscopic and leads, in general, to oddly curved tracks.\footnote{Note that the range $1~{\rm eV} \lesssim \Lambda \lesssim 10~{\rm keV}$ roughly corresponds to the length range $1~{\rm mm} \lesssim \ell_{\rm eff} \lesssim 100~{\rm km}$.  The distance $\sim 100~{\rm km}$ is still relevant for the LHC because the sagitta of an LHC track is roughly $d_{\rm max}^2/R$ where $d_{\rm max} \sim 1~{\rm m}$ is the radius of the tracker.  Taking the sagitta to be comparable to the tracker resolution $10~{\rm \mu m}$, one finds that there is sensitivity up to $R \sim 100~{\rm km}$.}  Finally, for $\Lambda \lesssim 1~{\rm eV}$ the effective string length is megascopic and does not play a role in collider searches, leaving the quirks to appear as stable charged particles.

By oddly curved, we simply mean that the quirk tracks deviate from the standard helix that curves in the $xy$-plane.  Quirk tracks can exhibit a wide variety of strange behaviors, for example, curving outside the $xy$-plane, reversing direction, or passing through the same detector layer more than once.  Crucially, however, they can also closely approximate a standard helix trajectory (at least inside the detector volume).  The $v^2$ factor in Eq.~\eqref{eq:leff} allows for a range of $\ell_{\rm eff}$ values for a given $\Lambda$. \\

%%%%%%%%%%%%%%%%%%%%%%%%%%%%%%%%%%%%%%%%%%%%%%%%%%%%%%%%%%%%%%%%%%%%%%%%%
\noindent {\bf Reinterpreting Existing Searches ---}
As mentioned above, for a wide range of $\Lambda$ there is a non-zero probability that a quirk track would be reconstructed at the LHC.  When this happens the quirk will appear simply as a heavy stable (or long-lived) charged particle.  In a collider, such particles are found by looking for tracks with large deposits of energy and/or a long time of flight (as compared to muons).  When both tracks fail to be reconstructed monojet searches will have sensitivity, provided that the quirks have been produced in association with a sufficiently energetic jet (through either initial or final state radiation).  Monojet searches look for large missing transverse energy that results from a jet recoiling against undetected particles.

\begin{figure} [bt]
\begin{center}
\includegraphics[width=0.40\textwidth]{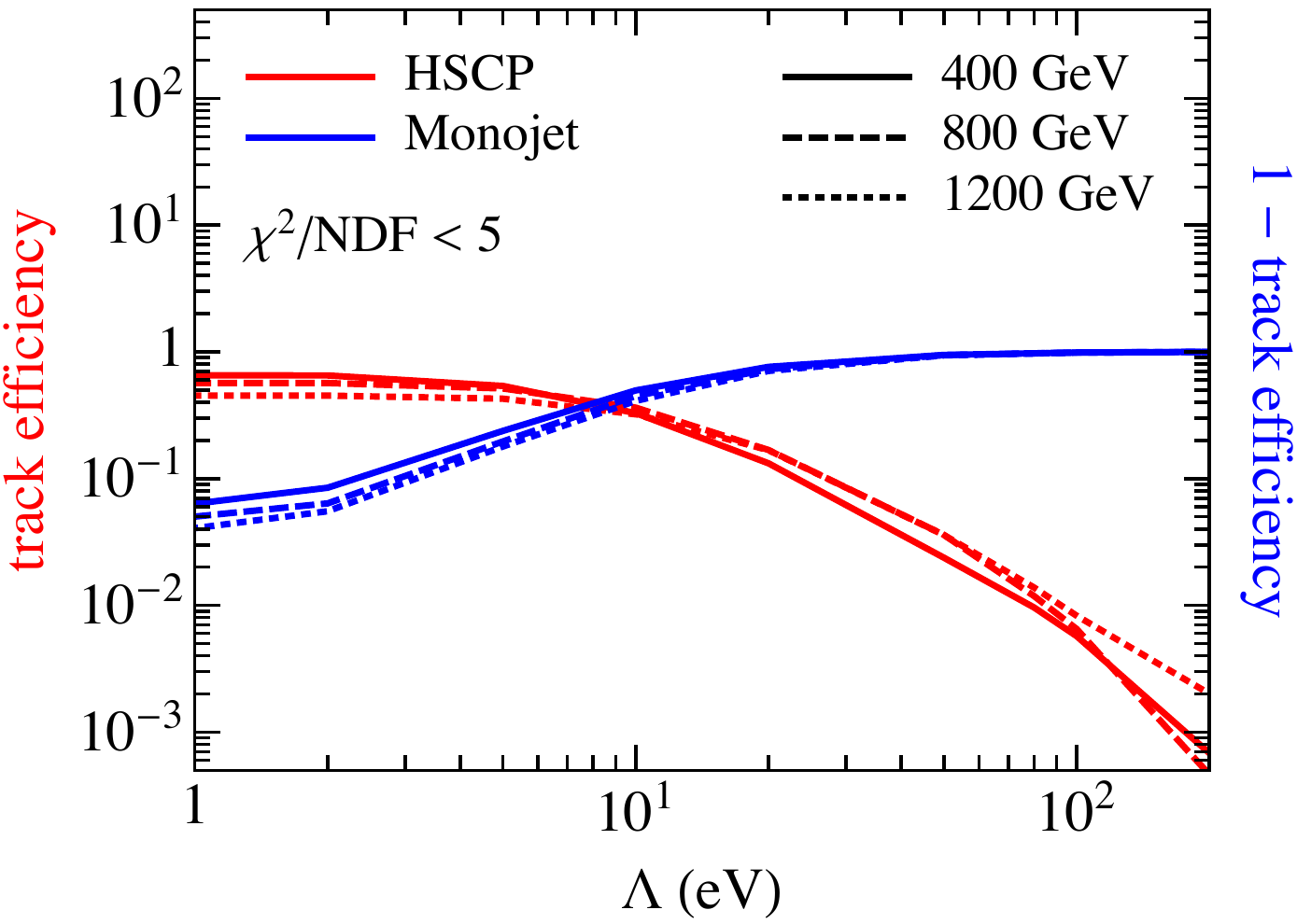}
\end{center}
\vspace{-.3cm}
\caption{Track efficiency as a function of confinement scale, $\Lambda$, for various quirk masses with a magnetic field of $\vec{B} = (0, 0, 3.8~{\rm T})$ at 13 TeV for HSCP events (red) and monojet events (blue).  The efficiencies do not asymptote to 0 or 1 at small $\Lambda$ due to the other track cuts applied (see Table~\ref{tab:cuts}).}
\label{fig:track_eff_lhc}
\end{figure}

The probability to reconstruct a track is characterized by the track efficiency and is shown in Fig.~\ref{fig:track_eff_lhc} for HSCP searches (red) and for monojet searches (blue).  The track efficiency is computed by applying a series of track selection cuts (as used by CMS in their HSCP analysis~\cite{CMS-PAS-EXO-16-036}) listed in Table~\ref{tab:cuts} and described below.

%%%%%%%%%%%
%%%%%%%%%%%

First, the quirk propagation is found by solving Eq.~\eqref{eq:eqm}.  We use the straight string approximation throughout.  Each time a quirk passes through a tracker layer it registers as a hit with an efficiency that we assume to be 100\%.  To account for the fact that in practice the hit efficiency is less than 100\% we increase the $n_{\rm hits}$ requirement to $\geq 9$ from the typical $\geq 8$.  The $n_{\rm hits}$ requirement specifies the minimum number of layers a track must hit.

We model the tracker geometry following the specifications of the CMS tracker,\footnote{We use the CMS tracker to maximize the accuracy of our results for the HSCP and 0T searches.} which consists of a combination of barrel layers and endcap layers that cover the range $-2.5~<~\eta~<~2.5$ where a particle will pass through $\approx~11-16$ layers depending on its trajectory~\cite{Chatrchyan:2008aa}.  Details of our tracker model are given in the appendix.

The tracks are then fit to a helix, which corresponds to the trajectory of a charged particle in the longitudinal magnetic field of the detector.  A helix is given by
\begin{subequations} \label{eq:helix} \begin{align}
h_x(t; R,\phi,\lambda) &= R \cos(\phi \pm (t/R)\cos\lambda) - \cos\phi, \\
h_y(t; R,\phi,\lambda) &= R \sin(\phi \pm (t/R)\cos\lambda) - \sin\phi, \\
h_z(t; R,\phi,\lambda) &= t\sin\lambda,
\end{align}\end{subequations}
where $t$ is the parameter along the curve, $R$ is the radius, $\phi$ is the initial azimuthal direction, and $\lambda$ is the dip angle.  For a completely general helix there are 3 additional parameters specifying the initial position of the helix, but we set this to the origin for simplicity.  The $\pm$ depends on the charge of the particle.

Next, we perform a $\chi^2/{\rm NDF}$ fit to Eq.~\eqref{eq:helix} and assume that quirks with $\chi^2/{\rm NDF}<5$ would be reconstructed as tracks~\cite{CMS-PAS-EXO-16-036}.  We use a spatial resolution of $30~{\rm \mu m}$ for each hit~\cite{Chatrchyan:2008aa}.  The $p_T^{\rm eff}$ cut is applied to the measured $p_T$ of the track (computed from the best fit $R$ value) rather than the true $p_T$ of the particle.

%%%%%%%%%%%%%%
\begin{table}
\begin{center}\begin{tabular}{c|ccc}

cut                 & HSCP       & monojet   & 0T        \\ \hline
$|\eta|$            & $<2.1$     & $<2.5$    & $<2.1$    \\
$n_{\rm hits}$      & $\geq 9$   & $\geq 9$  & $\geq 9$  \\
$v$                 & $>0.6$     & $-$       & $-$       \\
$p_T^{\rm eff}$     & $>65$ GeV  & $>10$ GeV & $>65$ GeV \\
$\chi^2/{\rm NDF}$  & $<5$       & $<5$      & $<5$      \\
$R$                 & $-$        & $-$       & $<1500$ m   
\end{tabular}
\end{center}
\caption{Cut flow for identifying a track.}
\label{tab:cuts}
\end{table}
%%%%%%%%%%%%%%

The first search that we reinterpret is the HSCP search.  The most sensitive HSCP searches have been performed at 13 TeV by CMS with 12.9~fb$^{-1}$~\cite{CMS-PAS-EXO-16-036} and by ATLAS with 3.2~fb$^{-1}$~\cite{Aaboud:2016uth} and are presented as mass limits on stable particles of different charges and SU(3) representations.  We follow the event selection of the former analysis which primarily consists of a $\chi^2/{\rm NDF}$ cut on the track and a cut of $p_T^{\rm eff}~>~65~{\rm GeV}$.  We only consider the sample that would be selected by the muon trigger which adds the additional requirement that $v~>~0.6$ in order that the trigger be $\approx 100\%$ efficient.  We generate quirk pair events using Madgraph 5 v2.3.3~\cite{Alwall:2014hca}.  The tracking efficiency is shown in Fig.~\ref{fig:track_eff_lhc} (red).  We define the track efficiency as the number of tracks passing all track requirements divided by the number of tracks passing the $|\eta|$ cut.  As expected, the efficiency drops as the string length becomes comparable to the detector scale, while it approaches 100\% when the string tension is small compared to the Lorentz force.  The asymptotic value at low $\Lambda$ is determined by primarily by the $v~>~0.6$ cut.

The second search we reinterpret is the monojet search.  From CMS the strongest search is at 13 TeV and uses 12.9~fb$^{-1}$~\cite{CMS-PAS-EXO-16-037} and from ATLAS it is at 13 TeV using 3.2~fb$^{-1}$~\cite{Aaboud:2016tnv}.  While the CMS search has better reach, we use the ATLAS limits because they are presented as limits on compressed stop squarks which is kinematically more similar to quirk events than the setups in the CMS search.  We also show limits scaling the ATLAS result up to 12.9 fb~$^{-1}$.  We generate quirk pair events along with a radiated jet of $p_T~>~200~{\rm GeV}$ and follow the monojet event selection in~\cite{Aaboud:2016tnv}.  The track selection in this case still identifies when a good track would be selected, however, opposite to the HSCP case, this means the event would be rejected.  For this reason we plot the quantity $(1-\text{track efficiency})$ in Fig.~\ref{fig:track_eff_lhc} (blue).  Note that even for small $\Lambda$ the efficiency does not go to zero because some tracks still fail the selection criteria.

\begin{figure}[bt]
\begin{center}
\includegraphics[width=0.40\textwidth]{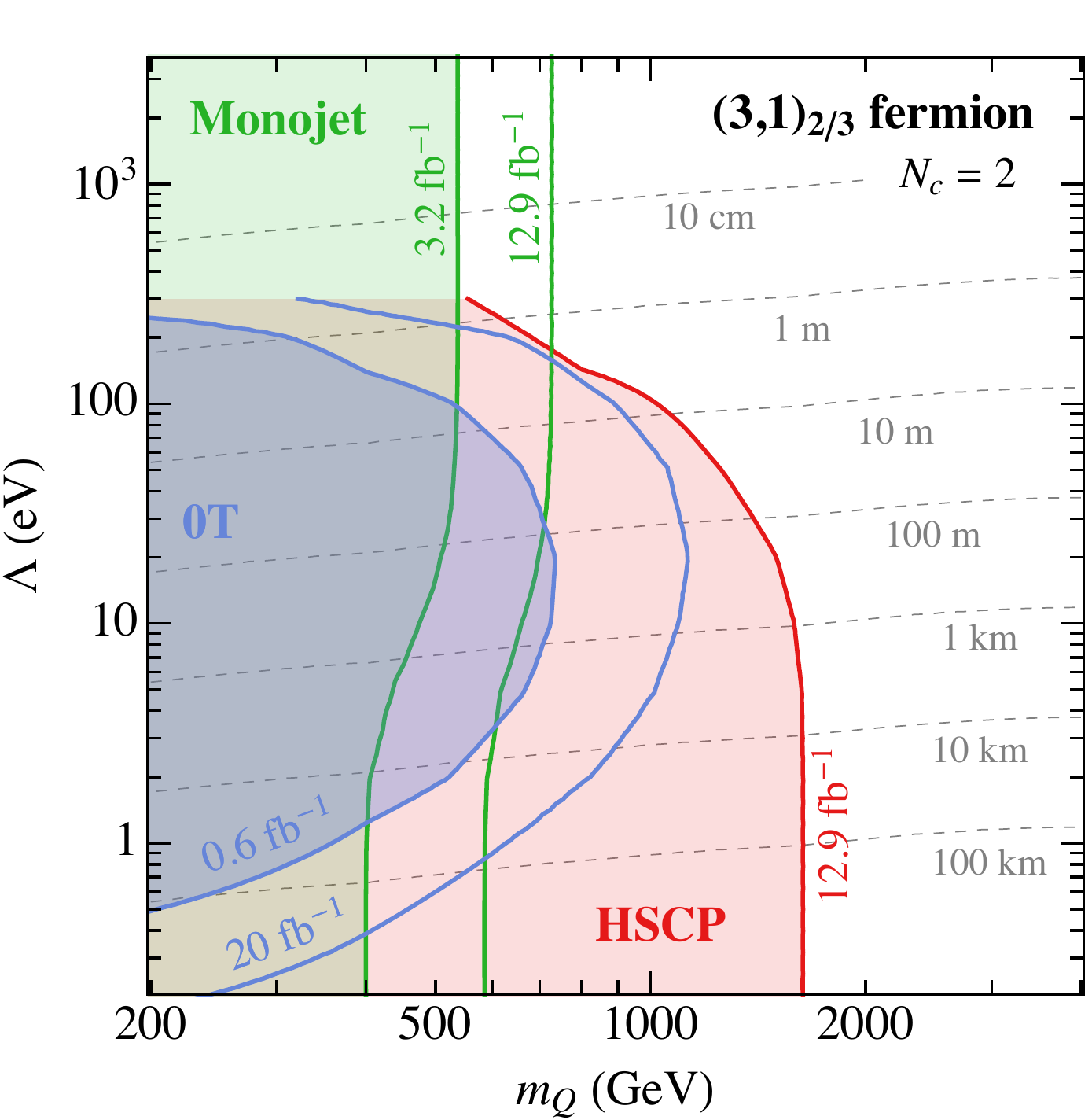}
\end{center}
\vspace{-.3cm}
\caption{The 95\% C.L. limits on a color triplet fermionic quirk with $N_c = 2$.  The red/green bound comes from HSCP/monojet searches and the blue bound is our projection for the 0T data.  The grey dashed lines show contours of $\ell_{\rm eff}$.}
\label{fig:limits_fermion_triplet}
\end{figure}

Using both HSCP and monojet searches provides a very complementary approach so that while our simplified tracker parametrization could differ from a full simulation it still captures the crucial features.  In particular, the complementarity ensures that our results reliably cover the full range of $\Lambda$ we study.  When one efficiency degrades, the other reaches its maximum (and at the maximum the tracking reconstruction details are less important).  Another important feature captured in Fig.~\ref{fig:track_eff_lhc} is the weak dependence on the quirk mass (since $\ell_{\rm eff}$ is approximately linear in mass).  In a similar manner, the quirk's electric charge has a minor impact on the efficiencies.  Indeed, while we use a fermionic color triplet quirk with quantum numbers $({\bf 3}, {\bf 1})_{2/3}$ under the SM gauge group as a case study, we will provide limits for a few other cases.   The results are shown in Fig.~\ref{fig:limits_fermion_triplet} which will be further explained in the following section. \\
 
%%%%%%%%%%%%%%%%%%%%%%%%%%%%%%%%%%%%%%%%%%%%%%%%%%%%%%%%%%%%%%%%%%%%%%%%%
\noindent {\bf Using the 0T Data ---}
In addition to the reinterpreted searches, we propose an entirely new search whose sensitivity is maximal in the challenging region near $\Lambda~\sim~10~{\rm eV}$.  This search makes use of the 0.6~fb$^{-1}$ of 13 TeV data with a 0T magnetic field~\cite{Khachatryan:2016hje}.  Without a magnetic field, all SM particles, both neutral and charged, travel in straight lines.  Quirks, on the other hand, still curve due to the string tension. This means one can simply count the number of curved tracks in the 0T data and accordingly set a limit or make an observation of quirks. Operationally, this would entail running the tracking algorithm on the 0T data while pretending there is a magnetic field and should not require any modifications to the tracking algorithm itself.

%%%%%%%%%%%
\begin{figure}[bt]
\begin{center}
\includegraphics[width=0.40\textwidth]{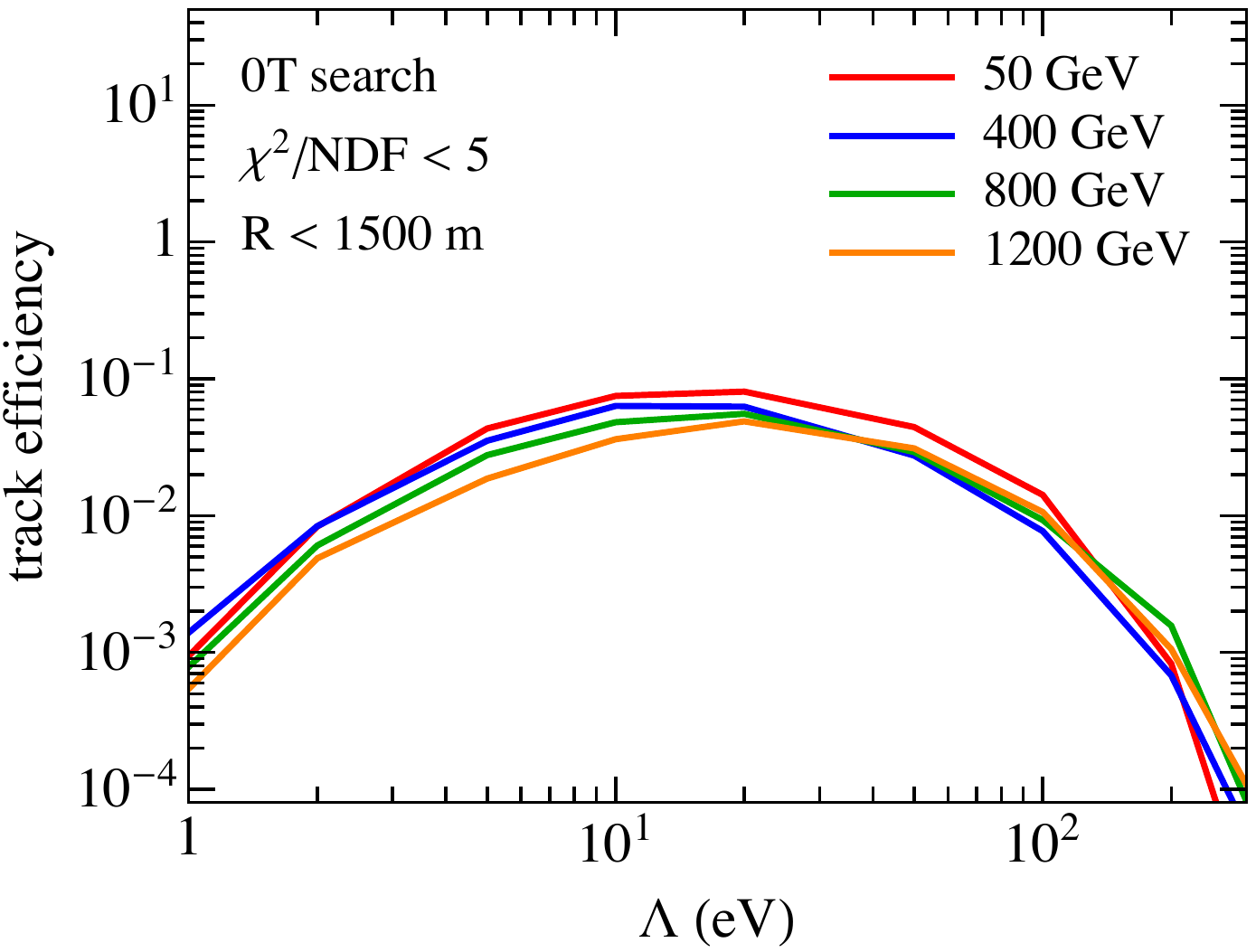}
\end{center}
\vspace{-.3cm}
\caption{Track efficiency as a function of confinement scale, $\Lambda$, for various quirk masses with a magnetic field of $\vec{B} = (0, 0, 0)$ at 13 TeV.}
\label{fig:track_eff_0T}
\end{figure}
%%%%%%%%%%%

The efficiency for identifying a track in the 0T data is shown in Fig.~\ref{fig:track_eff_0T}.  For the 0T search we use the monojet trigger~\cite{CMS-PAS-EXO-16-013} which uses an analysis-level cut of $p_T>100~{\rm GeV}$ for the leading jet and $\met>200~{\rm GeV}$ where $\met$ excludes muons (and would also exclude the quirk tracks).  While in principle one could use the muon trigger as in done in the HSCP searches, in practice the monojet trigger is more effective.  The reason is that without sizable initial radiation for the quirk system to recoil against, the quirks will be almost back-to-back in the transverse plane.  This means there is no curvature in the $xy$-plane so that the quirk trajectories cannot be reconstructed as non-straight helices.  Therefore, at least some initial state radiation is required for a non-zero efficiency.

We generate events with a single jet with $p_T~>~200~{\rm GeV}$ and apply the track cuts in Table~\ref{tab:cuts}.  These closely follow the selection from the HSCP search with the exception that we add a requirement that the fitted radius must be $R<1500~{\rm m}$.  We estimate the $R$ value from the sagitta $s$ of a track
\begin{equation}
s \approx \frac{d_{\rm max}^2}{8 R},
\end{equation}
where $d_{\rm max}$ is the chord length, corresponding to the radius of the tracker.  We take $d_{\rm max} \approx 1~{\rm m}$~\cite{Chatrchyan:2008aa} and set the sagitta to the single hit resolution of $s~\approx~30~{\rm \mu m}$~\cite{Chatrchyan:2008aa} and require a $3\sigma$ single hit fluctuation to find $R~\approx~1500~{\rm m}$.  Note that estimate assumes 3 hits.  Since a straight track faking a quirk requires $\geq 9$ hits, the fake rate is much lower than indicated by the $3\sigma$ requirement.  The $R$ cut is responsible for the decrease in track efficiency at $\Lambda~<~20~{\rm eV}$.

We assume that the fake rate is sufficiently low and that multiple scattering effects faking a curved track are sufficiently rare to treat the analysis as zero-background.  In principle, if a track with non-zero curvature is discovered the event could be inspected and checked for the presence of a second curved track, providing a smoking gun of the signal.  A limit is projected corresponding to observing $\leq 3$ events~\cite{Feldman:1997qc}.\\

%%%%%%%%%%%%%%%%%%%%%%%%%%%%%%%%%%%%%%%%%%%%%%%%%%%%%%%%%%%%%%%%%%%%%%%%%
\noindent {\bf  Discussion of Results ---}
The results are shown, for a $({\bf 3}, {\bf 1})_{2/3}$ fermion with $N_c=2$, in Fig.~\ref{fig:limits_fermion_triplet}.  The shaded red region shows the limits from HSCP searches which drive the limits for $\Lambda~\lesssim~200~{\rm eV}$.  The shaded green region shows the limits from the ATLAS monojet search that used 3.2~fb$^{-1}$ and the unshaded green limit scales up the limit to a dataset of 12.9~fb$^{-1}$ (the amount used in the CMS monojet search).  Our projection for the 0T search is given by the shaded blue region which uses 0.6~fb$^{-1}$.   The unshaded blue line shows a hypothetical dataset of 20~fb$^{-1}$ with no magnetic field and is the minimum amount of data required to probe parameter space beyond what is covered by HSCP and monojet searches.  The HSCP and 0T bounds are cut off at $\Lambda~=~300~{\rm eV}$ because our statistics there are insufficient for a reliable estimate.

Regarding QCD hadronization, the HSCP searches at the LHC use two different models~\cite{CMS-PAS-EXO-16-036,Aaboud:2016uth}.  The first model~\cite{Kraan:2004tz,Mackeprang:2006gx} assumes that the heavy hadrons can be charged or neutral when exiting the calorimeter while the second model~\cite{Mackeprang:2009ad} assumes the all heavy hadrons are neutral when exiting the calorimeter.  The 0T search only uses information from the tracker and therefore does not depend on this assumption.  We take the fraction of $R$-hadrons that are charged to be $0.55$ from Pythia 8~\cite{Sjostrand:2014zea}.

For the 0T and HSCP searches we allow for 1 or 2 identified tracks while for the monojet search we require 0 identified tracks.  The overall efficiency includes an acceptance factor ($\approx 85-95\%$ in the relevant region) that was not used in Figs.~\ref{fig:track_eff_lhc} and~\ref{fig:track_eff_0T}.  We find that HSCP can constrain higher masses than monojet searches or the 0T search.  This is not surprising as both the monojet and 0T searches require an additional radiated jet and have a larger background or smaller dataset.

In Table~\ref{tab:results} we report the limits for various other quantum numbers using $\Lambda~=~1~{\rm eV}$, $\Lambda~=~100~{\rm eV}$, and $\Lambda~=~10^3~{\rm eV}$ as benchmark points.

%%%%%%%%%%%
%%%%%%%%%%%

The gray lines in Fig.~\ref{fig:limits_fermion_triplet} show contours of constant $\ell_{\rm eff}$.  The $v$ used to compute $\ell_{\rm eff}$ is the mean of the velocity distribution at 13 TeV.  The $\ell_{\rm eff}$ contours give an idea of the length scales where each search is most effective and conversely, show where in parameter space other types of quirk searches would lie.

For $\Lambda \lesssim 1~{\rm eV}$, smaller than shown in Fig.~\ref{fig:limits_fermion_triplet}, the 0T limit quickly goes away but the HSCP limit stays constant at 1.6 TeV because $\ell_{\rm eff}$ simply gets larger.  On the other end, for $\Lambda \gtrsim 200~{\rm eV}$, the monojet limit stays constant until $\ell_{\rm eff} \sim {\rm \mu m}$ where the quirk system will appear as a single straight track.  Here, HSCP searches might have some sensitivity again.  For large enough $\Lambda$ eventually quirk-antiquirk annihilation can become prompt and searches for various resonances, like $\gamma\gamma$ or $j\gamma$ can be relevant~\cite{Kang:2008ea}.

Before concluding we note that when a quirk pair crosses, soft radiation could be emitted as hadrons or glueballs, leading to a loss of energy and modification of the quirk trajectories, as well as extra activity in the detector.  In most of the parameter space shown, however, we study $\ell_{\rm eff}~\gtrsim~1~{\rm m}$ so that the quirks typically do not cross each other.

Finally, one can ask how useful would modifications or extensions to tracking techniques be?  If it is possible to trigger on such anomalous tracks, one expects to set bounds competitive with HSCP searches and extend limits above $\Lambda~\gtrsim~200~{\rm eV}$.  For quirks, however, one limitation at the LHC is that the string length should be larger than the average distance traveled by the quirk between two layers of the tracker, which is $\approx~10~{\rm cm}$.  If not, then the sampling of the track would lose information on the peculiar ``periodic'' structure of the trajectory.  It is therefore in the region $10~{\rm cm}~\lesssim~\ell_{\rm eff}~\lesssim 10~{\rm m}$ where substantial improvements are possible.  This corresponds to an order of magnitude in $\Lambda$ and we believe that a more detailed study is warranted, not only at ATLAS and CMS, but also at more specialized detector such as MoEDAL~\cite{Acharya:2014nyr}.  \\
\begin{table}
\begin{center}
\begin{tabular}{ccc|ccc}
\multirow{2}{*}{spin}   & \multirow{2}{*}{charge} & \multirow{2}{*}{$N_c$} & $m_Q$ & $m_Q$ & $m_Q$ \\
        &                            &       & ($\Lambda =$ 1 eV)       &  ($\Lambda =$ 100 eV) &  ($\Lambda = 10^3$ eV) \\ \hline
fermion & $({\bf 3}, {\bf 1})_{2/3}$ &     2 & 1.6 TeV                  & 1.0 TeV               &  500 GeV               \\
scalar  & $({\bf 3}, {\bf 1})_{2/3}$ &     2 & 1.3 TeV                  & 700 GeV               &  350 GeV               \\
fermion & $({\bf 1}, {\bf 1})_{-1}$  &     2 & 650 GeV                  & 150 GeV               &  $-$                   \\
scalar  & $({\bf 1}, {\bf 1})_{-1}$  &     2 & 350 GeV                  &  60 GeV               &  $-$                   \\ \hline
fermion & $({\bf 3}, {\bf 1})_{2/3}$ &     5 & 1.8 TeV                  & 1.1 TeV               &  600 GeV               \\
scalar  & $({\bf 3}, {\bf 1})_{2/3}$ &     5 & 1.4 TeV                  & 850 GeV               &  450 GeV               \\
fermion & $({\bf 1}, {\bf 1})_{-1}$  &     5 & 800 GeV                  & 200 GeV               &   30 GeV               \\
scalar  & $({\bf 1}, {\bf 1})_{-1}$  &     5 & 450 GeV                  &  80 GeV               &  $-$                                     
\end{tabular}
\end{center}
\caption{Quirk mass limits for various quantum numbers at the benchmark points of $\Lambda=$~1~eV, $\Lambda=$~100~eV, and $\Lambda=10^3$~eV.}
\label{tab:results}
\end{table}

%%%%%%%%%%%%%%%%%%%%%%%%%%%%%%%%%%%%%%%%%%%%%%%%%%%%%%%%%%%%%%%%%%%%%%%%%
\noindent {\bf Outlook ---}
In this paper, we demonstrated that while quirk dynamics can result in very exotic tracks, they can also result in very standard looking tracks allowing for standard searches to constrain a substantial region of parameter space.  In particular, reinterpreting HSCP and monojet searches allows one to set limits in the regions $\Lambda~\lesssim~100~{\rm eV}$ and $\Lambda~\gtrsim~1~{\rm eV}$ respectively, which corresponds to effective string lengths of $\ell_{\rm eff}~\gtrsim~1~{\rm m}$ and $\ell~\lesssim~100~{\rm km}$.

For colored quirks the limits range from 1.0 TeV to 1.6 TeV with $N_c=2$.  The limits get correspondingly higher as $N_c$ is increased.  Limits on uncolored quirks were found to range from 150 GeV to 650 GeV.

We then proposed a novel use of the 0T data from CMS which involved looking for curved tracks in the dataset.  Given the small size of the dataset, 0.6~fb$^{-1}$, we predict that no curved tracks (at least due to quirks) should be observed in the data as HSCP searches already rule out the parameter space.  Amusingly, the 0T search could outdo the current HSCP limits if it had only $\gtrsim 20~{\rm fb}^{-1}$ of data.

We chose a few sample quantum numbers, in Table~\ref{tab:results}, but it would be interesting to see limits for a larger variety of quantum numbers.  On the experimental side, it would interesting to see if dedicated quirk searches can be done and would they compare to the monojet and HSCP searches.

Finally, we argued that in the region $10~{\rm cm}~\lesssim~\ell_{\rm eff}~\lesssim~10~{\rm m}$ there is an opportunity to go well beyond the sensitivity of current searches by developing novel tracking techniques.  Such techniques could then be applied to other physics cases, for instance, kinked or appearing tracks.  Given the simplicity of our 0T analysis, we hope that this work can serve as motivation for moving towards more involved tracking modifications in order to fully exploit the LHC's potential for unusual tracks.

%%%%%%%%%%%%%%%%%%%%%%%%%%%%%%%%%%%%%%%%%%%%%%%%%%%%%%%%%%%%%%%%%%%%%%%%%
\vspace{1em}
\begin{acknowledgements}
The authors would like to thank Raffaele Tito D'Agnolo, Markus Luty, Duccio Pappadopulo, Joshua Ruderman, Andreas Salzburg, and Kris Sigurdson for useful discussions and Yuri Gershtein, Philip Harris, Simon Knapen, Scott Thomas, and Nhan Tran for helpful discussions and reading the manuscript.  M.F. is supported in part by the DOE Grant DE-SC0010008.  M.L. is supported by a Frank and Peggy Taplin membership at the Institute for Advanced Study.  This work was partly completed at KITP, which is supported in part by the National Science Foundation under Grant No. NSF PHY11-25915 and at the Aspen Center for Physics, which is supported by National Science Foundation grant PHY-1066293.
\end{acknowledgements} 

%%%%%%%%%%%%%%%%%%%%%%%%%%%%%%%%%%%%%%%%%%%%%%%%%%%%%%%%%%%%%%%%%%%%%%%%%
\vspace{1em}
\begin{center}{\bf Appendix A: Tracker Model Details}\end{center}

We model the tracker following the specifications in~\cite{Chatrchyan:2008aa}.  The tracker is comprised of a cylindrical barrel that surrounds the beam pipe and two disks (end caps) on each side of the barrel.  We consider each layer to be of negligible thickness.  In cylindrical coordinates we specify barrel layers by their $r$ position and $z$ extent and disk layers by their $|z|$ position and $r$ extent.

The tracker consists of the pixel detector:
\begin{itemize}

%%%%%
\item {\bf Pixel barrel} (3 layers) \\
$r = 4.4~{\rm cm}, 7.3~{\rm cm}, 10.2~{\rm cm}$ \\
$|z| = 0~{\rm cm} - 26.5~{\rm cm}$

%%%%%
\item {\bf Pixel end cap} (2 layers) \\
$|z| = 34.5~{\rm cm}, 46.5~{\rm cm}$ \\
$r = 6~{\rm cm} - 15~{\rm cm}$

\end{itemize}

and the strip tracker:
\begin{itemize}

%%%%%
\item {\bf Tracker inner barrel} (4 layers) \\
$r = 25.5~{\rm cm}, 33.9~{\rm cm}, 41.85~{\rm cm}, 49.8~{\rm cm}$ \\
$|z| = 0~{\rm cm} - 70.0~{\rm cm}$

%%%%%
\item {\bf Tracker outer barrel} (6 layers) \\
$r = 60.8~{\rm cm}, 69.2~{\rm cm}, 78.0~{\rm cm}, 86.8~{\rm cm}, 96.5~{\rm cm}, 108.0~{\rm cm}$ \\
$|z| = 0~{\rm cm} - 109.0~{\rm cm}$

%%%%%
\item {\bf Tracker inner disk} (3 layers) \\
$|z| = 80.0~{\rm cm}, 85.0~{\rm cm}, 90.0~{\rm cm}$ \\
$r = 20.0~{\rm cm} - 50.0~{\rm cm}$

%%%%%
\item {\bf Tracker end caps} (9 layers) \\
$|z| = 124.0~{\rm cm}, 141.0~{\rm cm}, 155.0~{\rm cm}, 169.5~{\rm cm}, \\ 188.0~{\rm cm}, 207.5~{\rm cm}, 228.0~{\rm cm}, 253.5~{\rm cm}, 280.0~{\rm cm}$ \\
$r = 22.0~{\rm cm}, 22.0~{\rm cm}, 22.0~{\rm cm}, 30.0~{\rm cm}, \\ 30.0~{\rm cm}, 30.0~{\rm cm}, 42.0~{\rm cm}, 42.0~{\rm cm}, 52.0~{\rm cm}$  \\

\end{itemize}

\bibliography{refs}
\end{document}